%% file: main.tex
\documentclass[letterpaper, 9pt, twoside]{article}

\usepackage[english]{babel}

\usepackage[letterpaper,top=1.5cm,bottom=2cm,left=1.5cm,right=1.5cm,marginparwidth=1.75cm]{geometry}

\usepackage{amsmath}
\usepackage{float}
\usepackage{graphicx}
\usepackage{multicol}
\usepackage{enumitem}
\usepackage[colorlinks=true, allcolors=blue]{hyperref}
\usepackage{authblk}
\usepackage{booktabs}
\usepackage{caption}
\usepackage{fancyhdr}
\usepackage{placeins}  
\usepackage[textwidth=3.5cm]{todonotes}
\usepackage[backend=biber,style=nature,sorting=none]{biblatex}
\newcommand{\inserttitle}{Predicting Progression Events in Multiple Myeloma from Routine Blood Work}

\title{\textbf{\inserttitle}}

\author[1,2,3,4]{Maximilian Ferle$^*$}
\author[2,4]{Nora Grieb}
\author[3]{Markus Kreuz}
\author[4]{Uwe Platzbecker}
\author[1,2]{Thomas Neumuth}
\author[1,3,5]{Kristin Reiche}
\author[2]{Alexander Oeser$^\dag$}
\author[4]{Maximilian Merz$^\dag$}

\affil[1]{Center for Scalable Data Analytics and Artificial Intelligence (ScaDS.AI) Dresden/Leipzig, Universität Leipzig, Germany}
\affil[2]{Innovation Center Computer Assisted Surgery (ICCAS), University of Leipzig, Leipzig, Germany}
\affil[3]{Department of Diagnostics, Fraunhofer Institute for Cell Therapy and Immunology, Leipzig, Germany}
\affil[4]{Department of Hematology, Hemostaseology, Cellular Therapy and Infectiology, University Hospital of Leipzig, Leipzig, Germany}
\affil[5]{Institute for Clinical Immunology, University Hospital of Leipzig, Leipzig, Germany}
\date{}
\input{acronyms}

\begin{document}
    \maketitle
    \def\thefootnote{*}\footnotetext{Correspondence to: maximilian.ferle@uni-leipzig.de}\def\thefootnote{\arabic{footnote}}
    \def\thefootnote{\dag}\footnotetext{These authors contributed equally to this work and share last authorship}\def\thefootnote{\arabic{footnote}}

    \begin{abstract}
        \input{sections/abstract}
    \end{abstract}

    \begin{multicols}{2}

        \section{Introduction}\label{sec:introduction}
        \input{sections/body/introduction}

        \section{Results}\label{sec:results}
        \input{sections/body/results}

        \section{Discussion}\label{sec:discussion}
        \input{sections/body/discussion}

        \section{Methods}\label{sec:methods}
        \input{sections/body/methods}

        \section{Data availability}\label{sec:data_availability}
        \input{sections/declarations/data_availability}

        \section{Code availability}\label{sec:code_availability}
        \input{sections/declarations/code_availability}

        \section{Acknowledgements}\label{sec:acknowledgements}
        \input{sections/declarations/acknowledgements}

        \section{Author contributions}\label{sec:author_contributions}
        \input{sections/declarations/author_contributions}

        \section{Competing interests}\label{sec:competing_interests}
        \input{sections/declarations/competing_interests}

    \end{multicols}

    \printbibliography

    \pagebreak
    \input{sections/supplementals}

\end{document}

%% file: acronyms.tex
\usepackage{acro}
\DeclareAcronym{mmrf}{
    short=MMRF,
    long=Multiple Myeloma Research Foundation,
}
\DeclareAcronym{mm}{
    short=MM,
    long=Multiple Myeloma,
}
\DeclareAcronym{iss}{
    short=ISS,
    long=International Staging System,
}
\DeclareAcronym{ndmm}{
    short=NDMM,
    long=newly diagnosed multiple myeloma,
}
\DeclareAcronym{imwg}{
    short=IMWG,
    long=International Myeloma Working Group,
}
\DeclareAcronym{pd}{
    short=PD,
    long=progressive disease,
}

\DeclareAcronym{hb}{
    short=Hb,
    long=Hemoglobin,
}
\DeclareAcronym{ca}{
    short=Ca,
    long=Calcium,
}
\DeclareAcronym{cr}{
    short=Cr,
    long=Creatinine,
}
\DeclareAcronym{wbc}{
    short=WBC,
    long=White blood bells,
}
\DeclareAcronym{mpr}{
    short=M-Pr,
    long=M-Protein,
}
\DeclareAcronym{sfl}{
    short=SFL,
    long=serum free light-chain
}
\DeclareAcronym{sflk}{
    short=SFL-$\kappa$,
    long=serum free light-chain $\kappa$
}
\DeclareAcronym{sfll}{
    short=SFL-$\lambda$,
    long=serum free light-chain $\lambda$
}
\DeclareAcronym{ldh}{
    short=LDH,
    long=Lactate dehydrogenase,
}
\DeclareAcronym{alb}{
    short=Alb,
    long=Albumin,
}
\DeclareAcronym{b2m}{
    short=$\beta$2m,
    long=$\beta$-2-microglobulin
}

\DeclareAcronym{pdf}{
    short=PDF,
    long=probability density function,
}
\DeclareAcronym{locf}{
    short=LOCF,
    long=\textit{Last Observation Carried Forward},
}
\DeclareAcronym{auroc}{
    short=AUROC,
    long=area under the receiver operating characteristic curve,
}
\DeclareAcronym{prc}{
    short=PRC,
    long=precision-recall curve,
}
\DeclareAcronym{auprc}{
    short=AUPRC,
    long=area under the \ac{prc},
}
\DeclareAcronym{roc}{
    short=ROC,
    long=Receiver Operating Characteristic,
}
\DeclareAcronym{ppv}{
    short=PPV,
    long=positive predictive value,
}
\DeclareAcronym{vvuq}{
    short=VVUQ,
    long=Verification\, Validation and Uncertainty Quantification,
}

\DeclareAcronym{ml}{
    short=ML,
    long=machine learning,
}
\DeclareAcronym{nn}{
    short=NN,
    long=neural network,
}
\DeclareAcronym{lstm}{
    short=LSTM,
    long=\textit{Long Short-Term Memory},
}
\DeclareAcronym{crbm}{
    short=CRBM,
    long=\textit{Conditional Restricted Boltzmann Machine},
}
\DeclareAcronym{vht}{
    short=VHT,
    long=vitural human twin,
}
\DeclareAcronym{arima}{
    short=ARIMA,
    long=\textit{AutoRegressive Integrated Moving Average},
}

%% file: sections/abstract.tex
The ability to accurately predict disease progression is paramount for optimizing multiple myeloma patient care.
This study introduces a system designed to forecast disease trajectories to enable the prediction of progression events in affected patients participating in the CoMMpass study by the Multiple Myeloma Research Foundation.
Our system leverages a hybrid neural network architecture, combining Long Short-Term Memory networks with a Conditional Restricted Boltzmann Machine, to predict future blood work from a series of historical laboratory results.

We demonstrate that our model can replicate the statistical moments of the time series ($0.95~\pm~0.01~\geq~R^2~\geq~0.83~\pm~0.03$) and forecast future blood work features with high correlation to actual patient data ($0.92\pm0.02 \geq r \geq 0.52 \pm 0.09$).
Subsequently, a second Long Short-Term Memory network is employed to detect and annotate disease progression events within the forecasted blood work time series.
We show that these annotations enable the prediction of progression events with significant reliability (AUROC $= 0.88 \pm 0.01$), up to 12 months in advance (AUROC(t+12~mos)$=0.65 \pm 0.01$).

Our system is designed in a modular fashion, featuring separate entities for forecasting and progression event annotation.
This structure not only enhances interpretability but also facilitates the integration of additional modules to perform subsequent operations on the generated outputs.
Our approach utilizes a minimal set of routine blood work measurements, which avoids the need for expensive or resource-intensive tests and ensures accessibility of the system in clinical routine.
This capability allows for individualized risk assessment, supports clinicians in scheduling risk-adapted follow-ups and making informed treatment decisions tailored to a patient's unique disease kinetics.
The represented approach contributes to the development of a scalable and cost-effective virtual human twin system for optimized healthcare resource utilization and improved patient outcomes in multiple myeloma care.

%% file: sections/body/introduction.tex
\ac{mm} is a heterogeneous disease with survival ranging from months to decades based on a patient's individual risk profile\supercite{usmani_clinical_2018}.
Characteristics associated with outcomes include disease-specific and patient-derived factors, as well as treatment-associated features, like the emergence of side effects and access to latest treatment innovations.
In the past, several scoring systems have been implemented to estimate the prognosis of \ac{mm} patients with \ac{ndmm}.
While the initial version of the \ac{iss}\supercite{greipp_international_2005} accounted for factors associated with disease burden, the first (R-ISS)\supercite{palumbo_revised_2015} and second revisions (R2-ISS)\supercite{dagostino_second_2022} of the \ac{iss} integrated cytogenetic data and lactate dehydrogenase to appreciate disease biology and aggressiveness.
More lately, clinical, demographic, genomic, and therapeutic data\supercite{maura_genomic_2024} as well as immune signatures have been used to assign patients into different prognostic risk categories\supercite{maura_genomic_2023}.
However, the majority of those established scoring systems solely rely on the initial assessment of the patient.
Longitudinal changes upon treatment initiation, like deterioration of laboratory values or disease kinetics are usually not accounted for.
According to current guidelines, patients with \ac{mm} undergoing treatment for newly diagnosed or relapsed disease should be followed-up at least every 12 weeks\supercite{dimopoulos_multiple_2021}.
These routine visits should include a clinical assessment and laboratory testing with a complete blood count, renal/liver function testing as well as myeloma-specific markers of disease activity.
The regular monitoring and quarterly lab testing for \ac{mm} patients produces an extensive collection of longitudinal data, presenting a significant challenge for analysis with traditional methods for survival analysis.
This wealth of data creates a growing need for more advanced processing techniques that can streamline the integration of this data, facilitate the identification of trends, predict patient outcomes, and optimize individualized care plans.
In recent years, \ac{nn} based models, specifically \ac{lstm} networks\supercite{hochreiter_long_1997}, were shown to be effective in analyzing and modeling disease trajectories in a variety of cancer types, outperforming various other methods for time-series analysis.
Here, applications ranged from the detection of brain tumors based on sequential magnetic resonance images\supercite{amin_brain_2020}, predicting survival outcomes in prostate cancer from longitudinal clinicopathological data\supercite{koo_long_2020}, using time-series tumor marker data for early detection of several types of cancer\supercite{wu_long_2022} to predicting cancer symptom evolution on the basis of past routinely collected nursing documentation\supercite{chae_prediction_2024}.
However, the use of sophisticated \ac{ml} techniques to predict patient outcomes in \ac{mm} is still in its infancy.
A very recent review by Allegra et al.\supercite{allegra_machine_2022} summarized the latest advances in utilizing \ac{ml} techniques for diagnosis, prognosis, and treatment selection based on clinical and gene expression data.
Therein, it is described how random forests\supercite{mosquera_orgueira_survival_2021, borisov_machine_2021, venezian_povoa_machine_2021, mitra_gene_2017}, clustering and graph-based approaches\supercite{borisov_machine_2021, venezian_povoa_machine_2021, ubels_gene_2020, mitra_gene_2017} as well as \ac{nn}s\supercite{borisov_machine_2021, venezian_povoa_machine_2021} can predict survival outcomes based on treatment, clinical and expression data.
However, these applications rely on stationary (mostly initial) assessments and do not account for longitudinal kinetics.
To the best of our knowledge, no prior study investigated the potential of leveraging the wealth of longitudinal data that accumulate in routine \ac{mm} care.
In this study, we introduce a predictive framework for analyzing individual disease trajectories of \ac{mm} patients and, subsequently, the prediction of individual progression events.
Our study demonstrates how \ac{ml} can be used to individualize and periodically refine risk assessment based on existing routine laboratory values.
These findings could play a crucial role in dynamically refining follow-up schedules or introducing preventative treatments for patients who are at an impending risk of future disease progression.
This approach aligns with the vision of the \ac{vht} that facilitates the development and validation of patient-specific predictive models, thereby enhancing clinical decision support and personalized health forecasting\supercite{viceconti_position_2024}.

%% file: sections/body/results.tex
\begin{figure*}[t]
    \includegraphics[width=\linewidth]{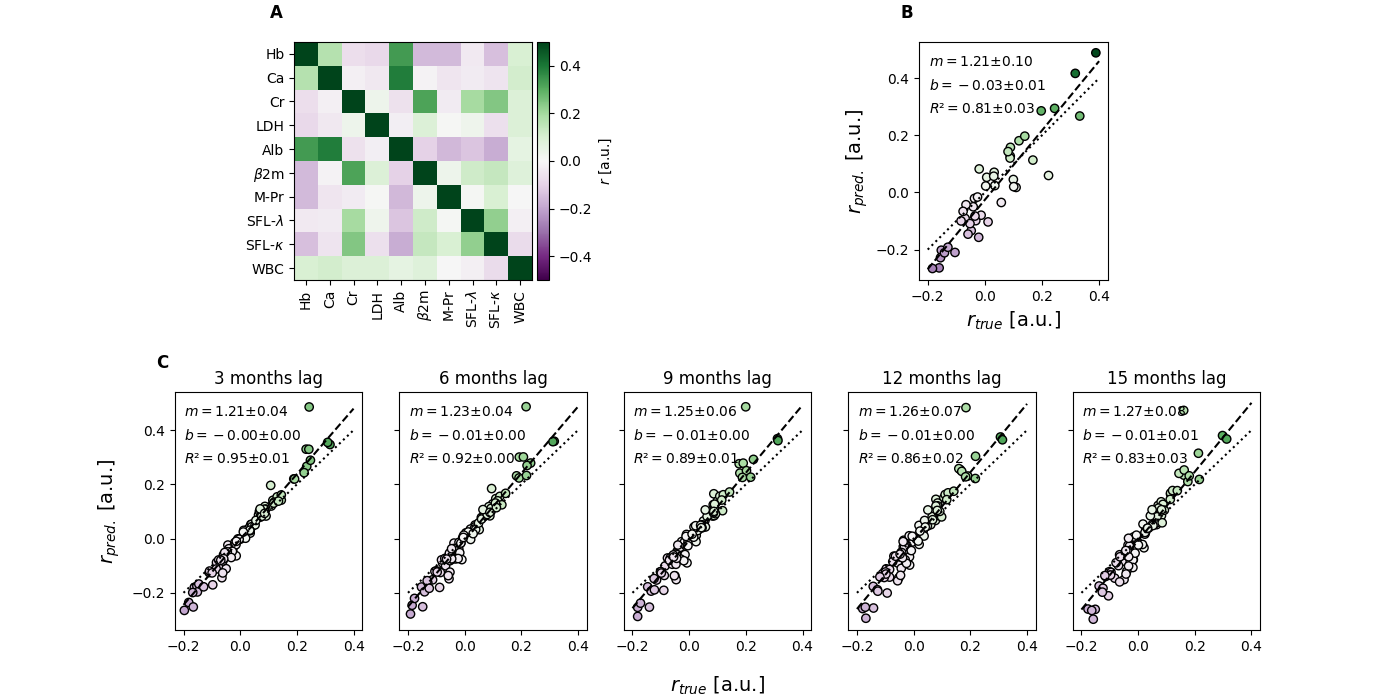}
    \caption{Statistical moments of the blood work time series data. (\textbf{A}) Correlation matrix of the ten chosen blood work parameters infered from the actual patient data. (\textbf{B}) Juxtaposition of the correlation coefficients from the actual data with those obtained from the forecasted data across varying forecasting horizons.
    Forecasting horizons were ranging from 3 to 15 months in 3-month increments and were summarized to reflect the overall performance across all forecasting horizons. (\textbf{C}) Correlation coefficients for each pair of features across five distinct lag times: 3, 6, 9, 12, and 15 months. Observed data is juxtaposed against forecasted data. Dashed lines show the lines of best fit. Parameters are reported as mean $\pm$ s.d. across cross-validation folds. Dotted lines show the expected lines of best fit for $m=1$ and $b=0$.}
    \label{fig:correlations}
\end{figure*}

\begin{figure*}[t]
    \includegraphics[width=\linewidth]{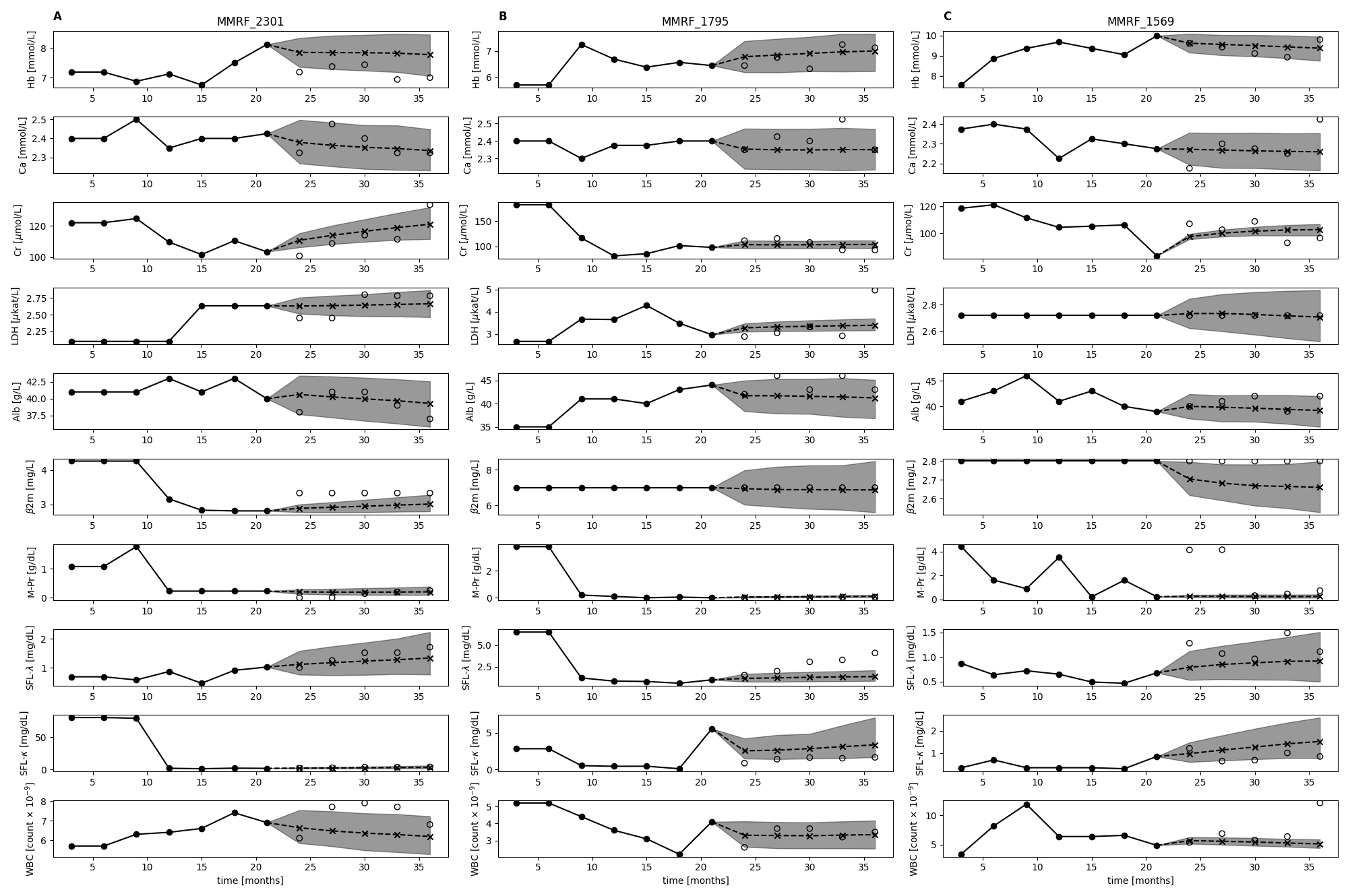}
    \caption{Forecasted patient trajectories of three individual patients undergoing Bortezomib- (\textbf{A}), Carfilzomib- (\textbf{B}) and IMIDs-based (\textbf{C}) treatments. The forecasting model was provided with the initial seven follow-ups, corresponding to 21 months of clinical data, to forecast the subsequent five follow-ups, covering an additional 15 months. Dashed lines and crosses show forecasts, circles show actual observations. Grey sleeves indicate the 95\% confidence interval of the distribution of forecasted trajectories.}
    \label{fig:patient_trajectories}
\end{figure*}

\begin{figure*}[t]
    \includegraphics[width=\linewidth]{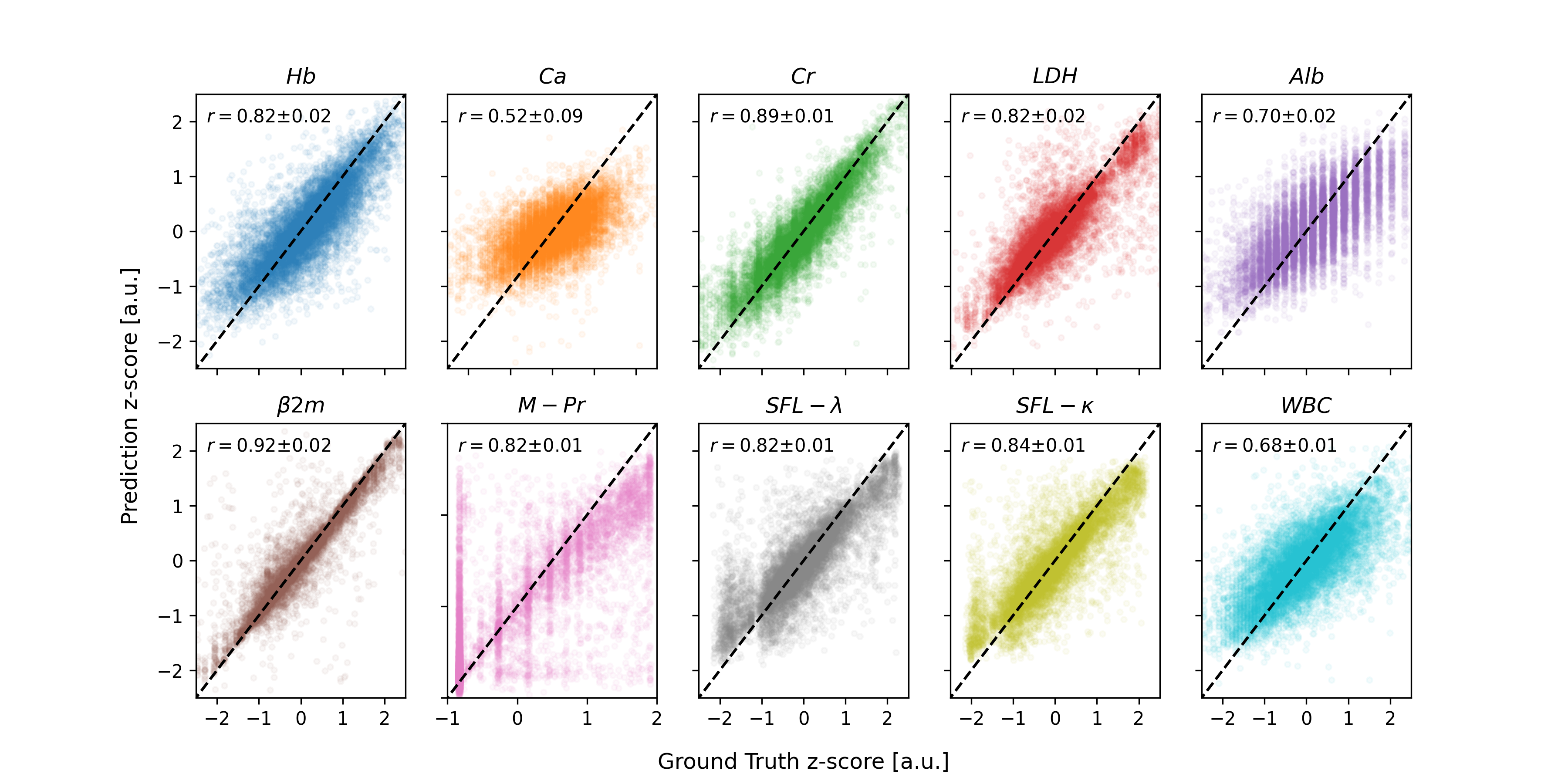}
    \caption{Juxtaposition of observed and forecasted blood work parameters at the next three-month follow-up. Values are shown in power-transformed state. Dashed lines indicate the trend lines, which values are expect to fall on. \textit{r} indicates Pearson's correlation coefficient. Correlation coefficients are reported as mean $\pm$ s.d. across cross-validation folds.}
    \label{fig:feature_predicitons}
\end{figure*}

\begin{figure*}[t]
    \centering{
        \resizebox{.8\linewidth}{!}{
            \includegraphics[width=\linewidth]{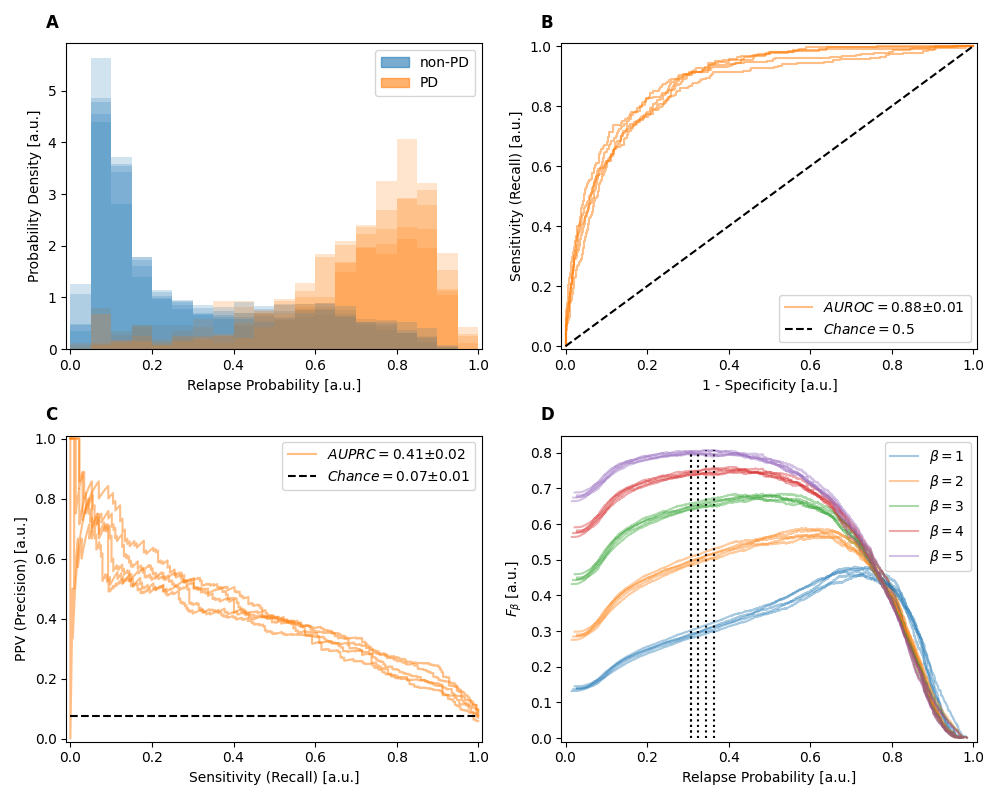}
        }}
    \caption{Annotation model performance data. (\textbf{A}) PDF for the output probabilities of the annotation model, color-coded for PD and non-PD instances. PDFs from the five cross-validation folds are overlaid with 20\% opacity to visually represent the consistency across models. (\textbf{B}) ROC curves of the annotation model across cross-validation folds. AUROC $= 0.88 \pm 0.01$. (\textbf{C}) PRC curves of the annotation model across cross-validation folds. AUPRC $= 0.41 \pm 0.02$. (\textbf{D}) $F_\beta$ curves for $\beta$ values of 1 through 5. Dotted lines indicate the relapse probability cut-off value of $0.33 \pm 0.02$ yielding an optimal $F_5$ value of $0.80 \pm 0.00$. All values are reported as mean $\pm$ s.d. across cross-validation folds.}
    \label{fig:baseline_roc}
\end{figure*}

\begin{figure*}[t]
    \includegraphics[width=\linewidth]{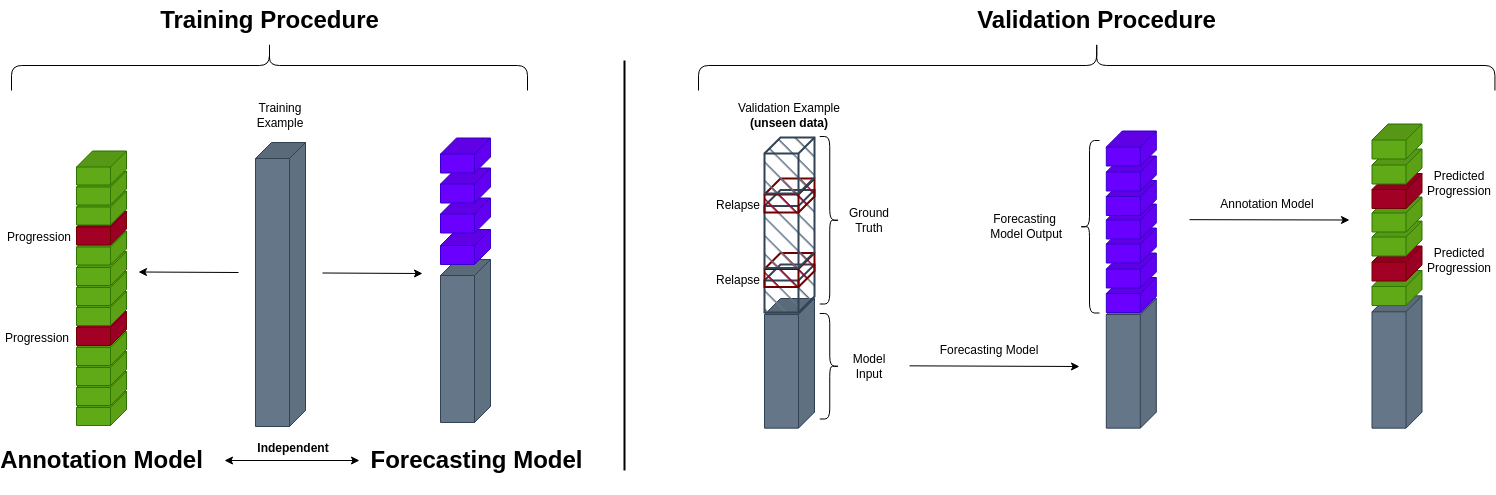}
    \caption{Model combination diagram illustrating the interplay between the forecasting and the annotation model. Both models were trained independently on the same training data to perform their respective tasks. During validation, the forecasting model was given the first \textit{n} follow-ups of a patient as input to forecast the next \textit{m} follow-ups. The annotation model was then tasked to flag progression events within the forecasted data. The thereby predicted progression events were evaluated against the actually observed future progression events in the down-stream analyses.}
    \label{fig:prediciton_diagram}
\end{figure*}

\begin{figure*}[t]
    \includegraphics[width=\linewidth]{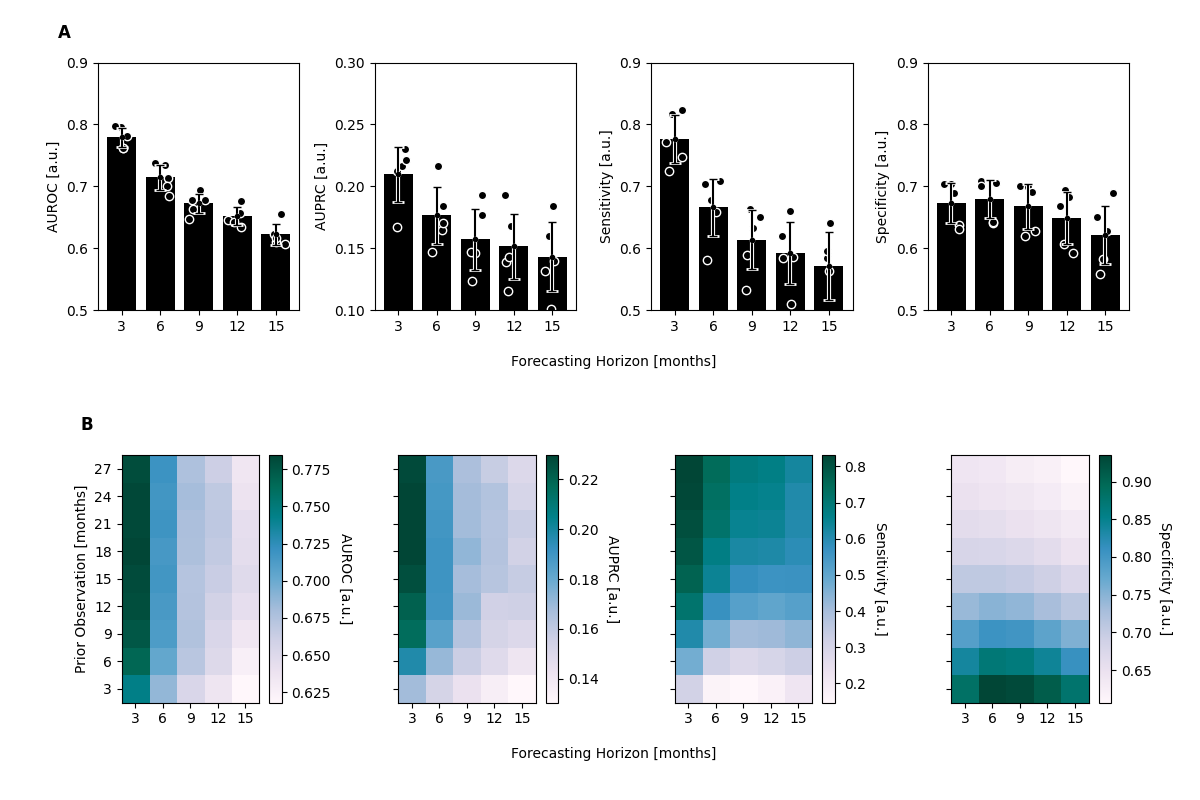}
    \caption{Evaluation of the forecasted progression events. (\textbf{A}) Overall AUROC, AUPRC, Sensitivity and Specificity across all time points for five distinct forecasting horizons: 3, 6, 9, 12, and 15 months. Bars and error bars indicate mean $\pm$ s.d. across the corss-validation folds. (\textbf{B}) Heatmaps indicating the AUROC, AUPRC, Sensiticvity and Specificity charted according to the amount of available prior patient observation (y-axis) and forecasting horizon (x-axis). Values are reported as means across cross-validation folds.}
    \label{fig:forecast_roc}
\end{figure*}

Our study aimed to develop and validate a model capable of generating likely future blood work for \ac{mm} patients based on the patients unique previous trajectories  (\textit{Forecasting Model}; see methods \ref{met:lstm_crbm} \& \ref{met:forecasting_model}).
Following this, we aimed to evaluate the predictive utility of these forecasts against a critical clinical endpoint to determine their effectiveness in anticipating such outcomes.
Given its significant clinical implications, \ac{pd} as defined by the \ac{imwg} consensus criteria\supercite{kumar_international_2016} was selected as the primary endpoint of interest.
To this end, we have implemented a second model to sufficiently identify and annotate progression events within \ac{mm} disease trajectories (\textit{Annotation Model}; see methods \ref{met:annotation_model}).
Consequently, we have applied the annotation model to the forecasted trajectories generated by the forecasting model to enable the anticipation of progression events well before they would otherwise become apparent.
The efficacy of the interplay of the two models was assessed through a series of analyses, the results of which are detailed below.

\subsection{Forecasting Model Performance}
Initial evaluations of the forecasting model revealed a high degree of accuracy ($R^2 = 0.81 \pm 0.03$) in capturing the correlations among the ten key features of the blood work derived from patient data  (see methods \ref{met:pat_charac}), as shown in the correlation matrix of figure \ref{fig:correlations}A.
Figure \ref{fig:correlations}B specifically focuses on comparing correlation coefficients.
It presents a scatterplot that juxtaposes the coefficients from the actual data with those obtained from the forecasted data across varying forecasting horizons.
Here, horizons ranged from 3 to 15 months, increasing in 3-month increments.
It is important to note that the model's ability to predict correlations accurately was strongest at the shortest horizon of three months and diminished with longer horizons.
For conciseness, this trend was aggregated in figure \ref{fig:correlations}B, to reflect the model's overall performance across all studied forecasting horizons.
Importantly, clinically relevant correlations like the positive correlation between serum \ac{cr} and \ac{b2m} levels or the inverse relationship between \ac{mpr} and \ac{alb} were reproduced.
In contrast, figure \ref{fig:correlations}C explores the temporal dynamics of the time series data by examining its lagged autocorrelation, a key statistical property.
This figure compares the lag-correlation coefficients for each pair of features in the observed data versus the forecasted data at five distinct lag times: 3, 6, 9, 12, and 15 months.
Each panel within figure \ref{fig:correlations}C provides a detailed view of the lag-correlations for each of the considered lag times.
The results demonstrate high accuracy in reproducing the autocorrelation of the time series, with $R^2$ values decreasing with increasing lag times: $R^2 = 0.95 \pm 0.01$ for a 3-months lag, $R^2 = 0.92 \pm 0.00$ for a 6-months lag, $R^2 = 0.89 \pm 0.01$ for a 9-months lag, $R^2 = 0.86 \pm 0.02$ for a 12-months lag, and $R^2 = 0.83 \pm 0.03$ for a 15-months lag.
These findings underscore the model's robustness in capturing the temporal dependencies between features, with a decrease in accuracy as the lag time increases.
In figure \ref{fig:patient_trajectories}, we present the predicted trajectories for three randomly selected patients undergoing Bortezomib- (Fig. \ref{fig:patient_trajectories}A) Carfilzomib- (Fig. \ref{fig:patient_trajectories}B) and IMIDs-based (Fig. \ref{fig:patient_trajectories}C) first-line treatments.
The forecasting model was provided with the initial seven follow-ups, corresponding to 21 months of clinical data, to forecast the subsequent five follow-ups, covering an additional 15 months.
Note that the forecasting model was not provided with information about the ongoing treatment and thus forecasted future blood work data merely based on the kinetics of the patients' unique trajectories.

The individual features in the model's prediction for the next three-month follow-up interval exhibited a strong correlation with the actual patient data (Fig. \ref{fig:feature_predicitons}), with coefficients ranging from $r = 0.52 \pm 0.09$ (Ca) to $r = 0.92\pm0.02$ ($\beta$2m).
We extended our analysis to examine the correlations of the blood work for predictions of multiple future follow-ups.
For multi-step predictions, the model's performance was observed to diminish as the forecasting horizon extended (table \ref{tab:abs_correlations}).
Nevertheless, the correlation between the features in the predicted and the actual blood work remained significantly above what would be expected by chance, with the smallest $r$-value recorded being $r = 0.34 \pm 0.08$ for Calcium at 15 months forecast interval.
Realizing that most features exhibit strong inertia, we investigated how well the model captures trends in the data by examining the differences of forecasted values to the last observed value (table \ref{tab:step_correlations}).
A lower, yet still moderate correlation was found ($0.62 \pm 0.04 \geq r \geq 0.25 \pm 0.03$) throughout all features in the blood work, indicating that the model was correctly forecasting the momentum of the data.
Here, the smallest r-value recorded was $r = 0.25 \pm 0.03$ for $\beta$2m at 6 and 9 months forecast interval respectively, while the largest $r$-value was $r = 0.62 \pm 0.04$ for Ca at 15 months forecast.
All reported correlation values were significant.

\subsection{Progression Annotation Model Performance}
In figure \ref{fig:baseline_roc}A, the \ac{pdf} for the output probabilities of the annotation model is displayed, color-coded for \ac{pd} and non-\ac{pd} instances.
The \ac{pdf}s from the five cross-validation folds are overlaid with 20\% opacity, allowing the combined opacity to visually represent the consistency across models.
The annotation model's initial performance was quantified using the \ac{auroc}, yielding a robust score of $0.88 \pm 0.01$ (Fig. \ref{fig:baseline_roc}B), indicating effective discrimination between \ac{pd} and non-\ac{pd} instances in the blood work time series data.
Given the low prevalence of progression events of $7\% \pm 1\%$ (see table \ref{tab:patient_table} and methods \ref{met:annotation_model}), we also considered the \ac{prc}, depicted in figure \ref{fig:baseline_roc}C.
The \ac{auprc} was found to be $0.41 \pm 0.02$, suggesting an average \ac{ppv} of 41\% across all potential decision thresholds.
To determine an optimal decision threshold while addressing the low prevalence issue, we optimized the $F_\beta$ score, which is defined as follows\supercite{baeza-yates_modern_1999}:
\[
    F_\beta = (1+\beta^2) \cdot \frac{\text{precision} \cdot \text{recall}}{(\beta^2 \cdot \text{precision}) + \text{recall}}
\]
Figure \ref{fig:baseline_roc}D presents the $F_\beta$ curves for $\beta$ values of 1 through 5.
We chose a $\beta$ of 5, thereby emphasizing recall over precision by a factor of five, leading to a decision threshold of $0.33 \pm 0.02$ and an $F_5$ of $0.80 \pm 0.00$.
The application of this optimized decision threshold to the testing data yielded a sensitivity of $0.92 \pm 0.02$ and a specificity of $0.65 \pm 0.03$ (Table \ref{tab:baseline_performance}).
Our choice for $\beta = 5$ was motivated by prioritizing higher sensitivity while accepting a higher false positive rate and lower specificity as a trade-off since missing a progression event would have more negative clinical relevance than indicating a progression event where there is none.
It is crucial to note that the $F_\beta$ optimization was conducted on the training data to avoid an overly optimistic bias in the model's performance when applied to the testing data.
\input{tables/baseline_performance_data}

\subsection{Progression Annotation in Forecasted Blood Work}
We examined the combined models' performance in predicting progression events depending on how far into the future the progression event would take place.
Figure \ref{fig:prediciton_diagram} conveys an overview of the interaction of the forecasting model and the annotation model.
In this scenario, the predictive power of the combination of both models started out high and showed a gradual decline as the forecasting horizon extended (Fig. \ref{fig:forecast_roc}A).
As detailed in figure \ref{fig:forecast_roc}A, the combined models' predictive performance, as measured by the \ac{auroc}, was $0.78 \pm 0.02$ at a 3-months forecasting horizon, diminishing to $0.62 \pm 0.02$ at 15 months.
The \ac{auprc} followed a similar trend, starting at $0.21 \pm 0.02$ for 3 months and decreasing to $0.14 \pm 0.03$ for 15 months.
Sensitivity started at $0.78 \pm 0.04$ and declined to $0.57 \pm 0.05$, while specificity slightly decreased from $0.67 \pm 0.03$ to $0.62 \pm 0.05$ over the same period.
These findings suggest that our models can predict progression events with high accuracy for a 3-months forecasting horizon and, with reliability that is still significantly beyond chance even for forecasting horizons extending beyond 12 months.

An additional layer of analysis was conducted to determine the impact of the amount of available prior patient observation on the accuracy of future forecasts.
Interestingly, the length of prior observation had little effect on short-term forecasting but became increasingly important for longer forecasting horizons.
This was visually represented by a color gradient in the heatmaps (Fig. \ref{fig:forecast_roc}B), with decreasing intensity from the upper left corner (long prior observation and short forecasting horizon) to the lower right corner (short prior observation and long forecasting horizon).
This trend was consistent across the \ac{auroc}, \ac{auprc} and Sensitivity.
For the \ac{auroc} heatmap, values are largely uniform across prior observation periods, with a slight tendency towards lower values at short prior observation periods.
Notably, the \ac{auroc} values fall within a range of approximately 0.62 to 0.78.
The \ac{auprc} heatmap appears to have a slightly more pronounced tendency to increase with larger prior observation periods.
The values are centered around a range of approximately 0.13 to 0.23.
Sensitivity, indicated by the third heatmap, presents the highest variance across different observation periods and forecasting horizons.
The values range from around 0.18 to 0.82, with a visually discernible patterning that suggests an increase in sensitivity associated with longer prior observation periods, indicating a stronger ability of the model to correctly identify progression events when provided with more historical data, especially when forecasting nearer-term events.
Lastly, the heatmap corresponding to specificity shows a broad spectrum of values as well, varying between approximately 0.63 to 0.92.
A counter-intuitive pattern emerged when examining the specificity, which was highest for short prior observations and longer forecasting horizons.
We saw that in scenarios, where there is little prior information available, the forecasting model engages in a quasi-random walk, devoid of meaningful information, leading the annotation model to predominantly label time points as non-\ac{pd}.
This tendency resulted in an artificially inflated specificity, rather than predictive performance of the model.
In summary, these heatmaps deliver a comprehensive overview of the model's performance, highlighting that while \ac{auroc} and \ac{auprc} remain relatively stable, Sensitivity and Specificity appear to be more sensitive to the duration of prior observation and the extent of the forecasting horizon.
This suggests that the length of prior observation, in conjunction with the forecasting horizon, plays a critical role in the model's ability to correctly predict impeding progression events.

%% file: tables/baseline_performance_data.tex
\begin{table}[H]
    \centering
    \caption{Annotation model baseline performance.}
    \begin{tabular}{c c c c}
        \toprule
        \textbf{AUROC}    & \textbf{AUPRC}  & \textbf{Sensitivity} & \textbf{Specificity} \\ 
        \hline
        $0.88 \pm 0.01$ & $0.41 \pm 0.02$ & $0.92 \pm 0.02$      & $0.65 \pm 0.03$      \\ 
        \bottomrule
    \end{tabular}
    \label{tab:baseline_performance}
\end{table}

%% file: sections/body/discussion.tex
In this study, we demonstrated how risk of impeding disease progression for individual \ac{mm} patients can be inferred from the kinetics in the unique trajectories of their routine blood work data.
For patients with \ac{mm}, the ability to anticipate future progression events is integral to effective patient care as this foresight allows for individualized risk assessment and the opportunity to adjust follow-up schedules and treatment plans proactively.
To this end, we employed a forecasting model to generate forecasted blood work data and let a second annotation model flag progression events within these forecasts.

The robustness of both our models was assessed using a 5-fold cross-validation to evaluate the models' performance across different subsets of the data.
The results of the cross-validation process are encouraging.
The models demonstrated a remarkable consistency in performance across all five folds, as evidenced by the very low variance in the relevant metrics.
Such consistency suggests that the models are not overly sensitive to specific partitioning of the data and can generalize well to new data.
Although, the \ac{mmrf} CoMMpass dataset represents one of the largest available, well-monitored collection of individual disease trajectories, one limitation of our study is the absence of external validation.
A dataset for external validation would ideally come from a separate origin, ensuring that it could provide a robust test of the models' predictive power in a truly independent context.
This would allow to assess the models' resilience to differences in data collection methods, demographic variations, and other external factors that could affect its performance.
Nevertheless, a major strength of using the \ac{mmrf} CoMMpass dataset to generate our virtual twin model is the wide range of state-of-the-art treatments used in the study as well as the heterogeneous patient population ranging from fit individuals, who are eligible for transplant, to frail patients treated with conventional therapies.
Because our model proved to be efficacious irrespective of the type of therapy and patients status, we are confident about the reproducibility in a wide range of clinical settings.
Future work should address this by securing access to external datasets for further validation.
Such efforts would not only enhance the credibility of the models but also potentially broaden their applicability, making them a more valuable tool in the field of \ac{mm}.

Another limitation that should be noted is that our model's current annotation of progression events exhibits a high rate of false positives as evident by the \ac{auprc}.
In our setup, if the model were to be implemented in a clinical setting to e.g. alert for risk of progression within the following three months, only about 20\% of such alerts would currently be justified.
However, it is important to contextualize this finding within the broader landscape of diagnostic testing in low-prevalence settings.
In scenarios where the actual incidence of a condition is rare, even tests with very high accuracy yield a large proportion of false positives\supercite{skittrall_specificity_2021, duijm_sensitivity_1997}.
This phenomenon is a consequence of the \ac{ppv} being inherently dependent on the prevalence of the condition in question\supercite{van_der_helm_application_1979, lu_channels_2020}. 
This is a well-documented challenge in medical diagnostics and affects a wide range of screening and monitoring tools\supercite{brown_interval_2003, lu_channels_2020, fox_common_2020}.
The clinical utility of a predictive model is thus not solely determined by its precision but also by the balance between the costs and benefits of false positives and false negatives.
Moreover, the deployment of such a model could be justified in settings where the cost of missing an actual progression event is substantially higher than the inconvenience or cost of additional follow-up.
The decision to implement such a model should be informed by a thorough cost-benefit analysis, considering the specific clinical context and the available resources.
In the case of our model, a false positive -- indicating a progression event where there is none -- would lead to earlier follow-up appointments, allowing for closer monitoring of at-risk patients rather than recourse to invasive diagnostics or treatments.
The models' high sensitivity when extensive historical patient data is available could be particularly beneficial in early detection and monitoring scenarios, provided that the healthcare system can accommodate the resultant need for increased follow-up.
Refinement of the model, particularly in conjunction with other diagnostic tools, could enhance its precision and expand its applicability in clinical practice even further.

Additionally, while our methods effectively quantify uncertainty within the confines of available data and predefined model structures, they do not fully account for systemic uncertainties that might arise from external factors such as changes in treatment protocols or new diagnostic criteria.
Our validation metrics, primarily \ac{auroc} and \ac{auprc}, while robust, do not capture all aspects of clinical utility, such as the cost-effectiveness of interventions based on model predictions or the impact of false positives on patient outcomes.
Furthermore, the dynamic nature of \ac{mm} and its treatment landscape means that the digital twin must continuously evolve, posing challenges for maintaining sustained model accuracy and relevance over time.
These limitations underscore the need for ongoing \ac{vvuq} processes and the exploration of more adaptive modeling techniques that can better accommodate the complexities and evolving nature of clinical environments.

The selection of an appropriate model for time series forecasting is pivotal to the success of any predictive analysis.
In this study, we opted to utilize a \ac{lstm} network over traditional methods such as \ac{arima} models.
This decision was informed by several factors that align with the unique characteristics of our data and the requirements of our predictive tasks.
Comparative studies have consistently shown that \ac{lstm}s outperform traditional methods, especially in scenarios where the data exhibit complex temporal dependencies\supercite{kontopoulou_review_2023, arunkumar_comparative_2022, kirbas_comparative_2020}.
The inherent structure of our data is multivariate, with several covariates influencing the target variables.
\ac{lstm}s are intrinsically designed to handle multivariate time series data, allowing them to capture the interdependencies among different covariates effectively\supercite{hochreiter_long_1997} rather than considering each covariate's trajectory in isolation.
This is a significant advantage over \ac{arima} and similar methods, which typically cannot natively accommodate multivariate time series\supercite{kontopoulou_review_2023}.
One of the most compelling reasons for choosing \ac{lstm}s over traditional methods is their feature learning capability.
This attribute eliminates the need for manual feature engineering, which can be time-consuming and may not capture all the predictive signals present in the data.
This allows \ac{lstm}s to learn from and adapt to the intricacies of the data, leading to more robust and generalizable models.

The ability to accurately forecast blood work holds significant clinical value, as it provides a window into the patient's future health status, allowing for preemptive medical interventions.
We have designed our prediction system as a modular pipeline, which features distinct entities for forecasting future blood work and for annotating progression events within these forecasts.
This modular approach facilitates interoperability and promises the possibility of integrating further modules for the annotation of alternative endpoints within the same system.
For instance, anticipating a drop in hemoglobin levels could prompt early iron supplementation or transfusions.
Similarly, forecasting a decline in white blood cell counts might lead to prophylactic measures to reduce the risk of infection before leukopenia sets in.
While disease progression was the primary endpoint we focused on in this study due to its clinical significance, the predictive modeling of blood work holds the potential to serve as a tool to anticipate a broader range of clinical endpoints, which would help guiding patient maintenance and thus enhance the quality of care.
This will be the scope of our future research endeavours.

Our model demonstrates a paradigm shift from classical prognostic risk evaluation based on initial assessments towards a framework in which a prediction of risk is made for specific future time points.
Moreover, the predictions of the models can be verified upon follow-up of the patient and its predictive capability can be periodically refined by updating predictions as soon as new patient data becomes available.
This interaction between patients and their computational representation allows to derive more informed treatment and health care decisions by e.g. shortening follow-up intervals for more close monitoring or early intervention at impeding risk of disease progression.
This foresight can be pivotal in clinical decision-making, enabling healthcare providers to implement timely therapeutic strategies to mitigate or prevent the adverse effects associated with such hematologic conditions.
Furthermore, our approach utilizes a minimal set of routine blood work measurements, avoiding the need for expensive or labor-intensive tests, ensuring practicality and wide-range accessibility in clinical settings.
In summary, our proposed system addresses several key challenges identified for the adoption of digital twins in healthcare\supercite{viceconti_position_2024}:
\begin{itemize}
    \item We exemplify the development of advanced, patient-specific predictive models (B1).
    \item By relying on routine blood work data, we address the need for representative, available data for model development and validation (B2).
    \item Our modular approach enhances interpretability and facilitates integration with additional modules, supporting the \ac{vht}'s emphasis on interoperability and scalability (B5).
\end{itemize}
By addressing these challenges, our system contributes to the realization of a scalable and cost-effective \ac{vht} ecosystem in \ac{mm} care, which could help improve patient outcomes and optimized healthcare resource utilization in \ac{mm}.


%% file: sections/body/methods.tex
\subsection{Patient Characteristics}\label{met:pat_charac}
Data from patients with \ac{ndmm} were retrieved from the CoMMpass study (NCT01454297) database version IA21.
To model \ac{mm} disease kinetics, we selected the following parameters for our predictive framework due to their respective importance in \ac{mm} disease monitoring:
\begin{itemize}
    \item \ac{hb} as anemia is a common complication in MM patients\supercite{cline_studies_1962, mittelman_implications_2003}.
    \item \ac{ca} due to its importance for bone turnover and hypercalcemia, which 30\% of MM patients present with\supercite{smith_multiple_2013, buege_corrected_2019}.
    \item \ac{cr} due to its involvement in the highly prevalent renal insufficiency of MM patients\supercite{knudsen_renal_1994, principal_discussant_winearls_acute_1995}
    \item \ac{wbc} counts, since leukopenia is a common side effect of anti-myeloma therapies and directly connected to infectious complications during therapy\supercite{stewart_carfilzomib_2015, rajkumar_lenalidomide_2010, palumbo_daratumumab_2016, dimopoulos_daratumumab_2016}.
    \item \ac{mpr}, \ac{sflk} and \ac{sfll} as \ac{mm} specific markers for disease activity\supercite{smith_multiple_2013}
    \item \ac{ldh}, \ac{alb} and \ac{b2m} due to their prognostic value for survival outcomes\supercite{palumbo_revised_2015}.
\end{itemize}
Furthermore, we have chosen the above parameters because they constitute routine measurements in \ac{mm} care and do not require costly or labor-intensive test and are therefore easily accessible in routine clinical practice if not already available.
Missing values were imputed through \ac{locf}.
If no value was available for \ac{locf}, we would carry the earliest available measurement backward.
Our rationale was to prevent cross-patient imputations and to mirror clinical practice as closely as possible.
The only patients excluded were those for which any of the before mentioned parameters were never measured across all their visits.
A total of 875 patients remained with an average of $19 \pm 9$ (mean $\pm$ s.d.) visits per patient.
We split this group into five distinct sets of 175 patients each for cross-validation purposes.
Thus, in each of the five folds, there were a total of 700 patients with $13{,}420 \pm 87$ unique time points available for training and 175 patients with $3{,}355 \pm 87$ unique time points for validation.
Out of these, $989 \pm 30$ time points marked progression events for training and $247 \pm 30$ for validation.
In table \ref{tab:patient_table} the relevant patient characteristics and distribution of data across the five folds are displayed.
\input{tables/patient_table}

\subsection{Data preprocessing}
All blood work parameters used in this study were subject to a feature-wise power transformation using the PowerTransformer class of scikit-learn\supercite{pedregosa_scikit-learn_2011} (version 1.4.1.post1) to approximate a Gaussian-like distribution of the data.
The PowerTransformer was fitted exclusively on the respective training portions of the data (80\%) and only applied to the validation portions (20\%) for transformation.
Figure \ref{fig:qq} contains QQ-plots of the distributions of the training and validation data after power transformation.

\subsection{Long Short-Term Memory Conditional Restricted Boltzmann Machine}\label{met:lstm_crbm}
Disease trajectories in \ac{mm} present a unique challenge due to their high heterogeneity and the presence of temporal patterns that operate over variable time scales\supercite{kumar_international_2016}.
The predictive modeling of disease trajectories in \ac{mm} requires the analysis of these temporal patterns to understand how they evolve over time.
Recognizing these complexities, we have implemented a hybrid model that utilizes a \ac{lstm} network in conjunction with a \ac{crbm}, a class of generative stochastic artificial neural networks\supercite{mnih_conditional_2012}.
\ac{crbm}s are a variant of the Restricted Boltzmann Machine\supercite{ackley_learning_1985}, which extend their capabilities by incorporating conditional dependencies on external or previous information.
In the context of disease trajectory modeling, this allows the model to learn the distribution of data conditioned on past information, making it particularly suitable for time-series data where the future state is dependent on observed historical data.
Because the outputs of \ac{crbm}s are inherently probabilistic, they facilitate distributions of potential disease trajectories, rather than yielding a single, deterministic forecast devoid of any quantification of uncertainty.
Such a probabilistic approach is particularly advantageous in medical applications where the exact future cannot be determined.
Hence, \ac{crbm}s were shown to be effective in modeling trajectories of diseases such as in Alzheimer's\supercite{fisher_machine_2019} and multiple sclerosis\supercite{walsh_generating_2020}.
However, the complexity of temporal dependencies that \ac{crbm}s can capture hinges on the number of previous time-lagged steps a prediction is conditioned on -- a design choice made when implementing the model\supercite{fisher_machine_2019}.
\ac{lstm}s address this challenge with their architecture: They are composed of modules with gating mechanisms -- the input, output, and forget gates -- that regulate information flow.
This allows the network to learn to retain relevant information and to discard the non-essential\supercite{hochreiter_long_1997}.
This selective memory enables the model to adaptively focus on the most predictive features over time.
Subsequently, the \ac{crbm} takes on the role of generating future time steps, conditioned on the internal state of the \ac{lstm}.
This approach ensures that predictions are always anchored to the full spectrum of available observations, rather than being limited to the most recent ones.

In summary, our design choice was motivated by the following rationale:
\begin{enumerate}
    \item The \ac{lstm} serves as a tool to encode the entire history of a patient's sequential blood work, effectively summarizing all prior observations within its internal state.
    This encoding is crucial for capturing the multi-scaled temporal patterns inherent in \ac{mm}.
    \item The integration of the CRBM facilitates a paradigm shift from traditional regression-based forecasting to probabilistic modeling.
    By generating probabilities conditioned on the encoded patient history from the \ac{lstm}, the \ac{crbm} provides a spectrum of possible outcomes rather than a single point prediction, thereby offering a more comprehensive understanding of uncertainty in \ac{mm} disease trajectories.
\end{enumerate}

\begin{figure*}[t]
    \includegraphics[width=\linewidth]{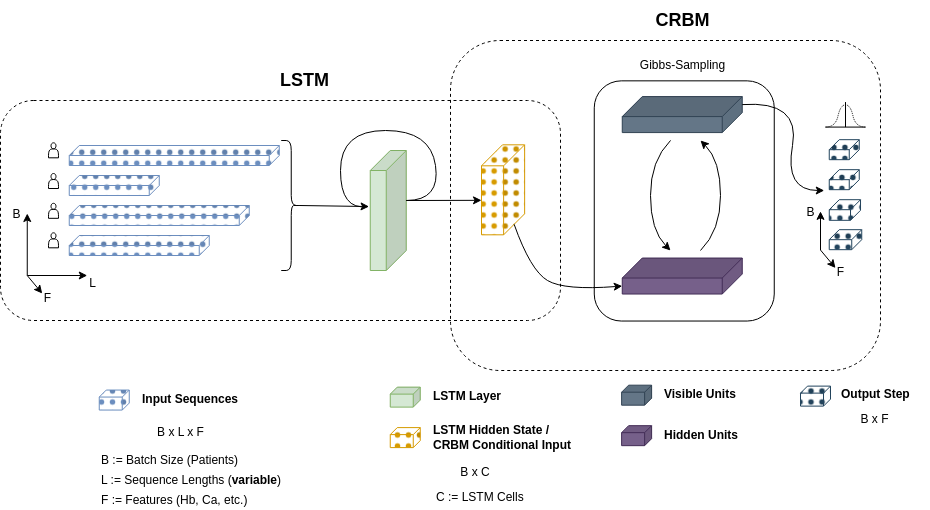}
    \caption{Architecture diagram of the forecasting model. The LSTM summarizes all observations of a patient’s sequential blood work within its internal state. The CRBM then generates probabilites of possible outcomes conditioned on the patient history encoded by the LSTM through a Gibbs-sampling algorithm.
    }
    \label{fig:lstm_crbm}
\end{figure*}

Our hybrid model, therefore, offers a more contextually aware framework to address the complex and variable temporal dynamics of \ac{mm}.
Figure \ref{fig:lstm_crbm} provides an overview of the model architecture and its internal information flow.
The \ac{lstm}-\ac{crbm} hybrid model is referred to as the \textit{Forecasting Model}.

\subsection{Model Implementation and Training}
All employed models were implemented using the PyTorch\supercite{paszke_pytorch_2019} framework (version 2.2.0) and trained using PyTorchLightning\supercite{falcon_pytorch_2023} (version 1.9.5).


\subsubsection{Forecasting Model}\label{met:forecasting_model}
The \ac{lstm} component of the forecasting model was configured with 10 input features which we selected as described in \ref{met:pat_charac} and consisted of 32 memory cells.
The \ac{crbm} component was greatly inspired by the Neural Boltzmann Machine architecture proposed by Lang et al.\supercite{lang_neural_2023}, with a significant adaptation that simplified the complexity of the model; We removed the latent spaces in the Bias-, Precision-, and Weights-networks, which decreased its computational complexity and potential for overfitting.
The \ac{crbm} was subsequently structured with 32 features in its input layer and 16 hidden units, which are connected to 10 visible units.
Total parameters: $11{,}572$.

Blood work was forecasted according to methods section \ref{met:monte_carlo}.

\subsubsection{Annotation Model}\label{met:annotation_model}
Given the critical importance of timely recognition of progression events in the management of \ac{mm}, we were motivated to develop an annotation model able to identify and flag such events within blood work time series data.
We pursued the development of this model with the goal of applying it to the forecasted blood work data to identify progression events well before they would otherwise become apparent.
The deployment of the model served two pivotal functions:
\begin{enumerate}
    \item Mitigating biases inherent in manual annotations and
    \item addressing the absence of certain features in routine tests, which are typically employed to annotate remission criteria according to the \ac{imwg}\supercite{kumar_international_2016}.
\end{enumerate}

To annotate a specific time point, the time series comprising all previous observations up to and including the current one were used as input for the annotation model.
Based on this, the model was trained to label the current time point as either \ac{pd} or non-\ac{pd}.
Labels were inferred from treatment response documentation in the CoMMpass dataset.
We observed that the prevalence of progression events across all time points in the data set was $7\% \pm 1\%$.
Therefore, to ensure a robust training process, we balanced the dataset by upsampling the \ac{pd} time series instances, ensuring the model was trained on an equal representation of both categories.
Testing was always done on the representative, imbalanced data.

The annotation model only utilized \ac{mpr}, \ac{sflk}, \ac{sflk}, as well as the SFL ratio and the SFL differences as input features.
Based on this, the Annotation Model employs an \ac{lstm} layer with 6 input features and 8 memory cells.
Following the \ac{lstm}, a fully connected dense layer with 8 input features and 2 output features is applied.
The dense layer utilizes a softmax activation function to output a probability distribution over the two annotation labels.
Total parameters: 530.

\subsubsection{Model Training}
The Forecasting model was trained in 100 Epochs with a batch size of 32 using the following weight decay parameters for the individual components:
\begin{itemize}
    \item LSTM: 0.1
    \item Precision-Net: 0.1
    \item Bias-Net: 0.1
    \item Weights-Net: 0.2
\end{itemize}

The Annotation model was trained in 200 Epochs with a batch size of 128 using the following weight decay parameters for its components:
\begin{itemize}
    \item LSTM: 1
    \item Dense: 1
\end{itemize}

Model hyperparameters and training parameters were determined through grid search.
The forecasting model was trained with \textit{Contrastive Divergence} loss and the annotation model was trained with \textit{Binary Cross-Entropy} loss.
Both models were trained with a learning rate of 0.0001 using the AdamW optimizer\supercite{loshchilov_decoupled_2019}.

\subsection{Monte Carlo Simulation}\label{met:monte_carlo}
The \ac{crbm} component of our forecasting model generates predictions by virtue of a Gibbs-sampling algorithm.
For an in depth explanation of the sampling algorithm, we would like to refer the reader to the paper by Lang et al.\supercite{lang_neural_2023} about the original implementation of a Neural Boltzmann Machine which we adopted for our model as described in section \ref{met:forecasting_model}.
In brief, to generate a prediction for a given time point, we drew $1{,}000$ samples employing 32 Markov-Chain steps per sample using all previous observations leading up to the one in question as input for our model.
We obtained the most probable estimate for each respective time point by averaging the resulting distribution of samples.
For multi-step predictions, we were drawing several samples in a recurrent loop, i.e. sample the immediate next step, reinserting it into the model, and then sampling successively until the desired forecasting horizon was reached.
This process enabled us to generate distributions of complete trajectories rather than isolated time points.
In coherence with single-step predictions, $1{,}000$ trajectories were sampled and averaged to yield the most probable trajectory for a given patient.

\subsection{Verification\, Validation and Uncertainty Quantification}
In accordance with best practices for \ac{vvuq} of digital twin systems, as outlined in the National Research Council's report \textit{Assessing the Reliability of Complex Models}\supercite{noauthor_assessing_2012}, our methodology addresses the components of \ac{vvuq} as follows;
\textit{Verification} was conducted by addressing to which degree the forecasting model's predictions replicated actual patient data, ensuring that the model accurately forecasts blood work values and effectively reflects changes between sequential measurements and maintains the integrity of cross-correlations between features and their lagged autocorrelation.
To this end, we utilized linear regression to obtain the slope, intercept and coefficient of determintation (\textit{$R^2$}) to assess the accuracy of the predicted cross-correlations between features, as well as Pearson's correlation coefficient (\textit{r}) to assess the correlation of changes in the individual features in the forecasted blood work and actual patient data.
\textit{Validation} was achieved through statistical measures; the \ac{auroc} and the \ac{auprc} were utilized to assess the discrimination between progression and non-progression time points and to evaluate the predictiveness of positive labels, respectively.
The combined efficacy of the forecasting- and annotation model was further validated by calibrating an optimal decision boundary and calculating the resulting sensitivity and specificity of the models' labelling.
\textit{Uncertainty Quantification} was addressed by assessing the impact of varying amounts of historical patient data and different forecasting horizons on the models' performance, measured by \ac{auroc}, \ac{auprc}, sensitivity, and specificity across different configurations.
Furthermore, we implemented a 5-fold cross-validation approach, allowing us to quantify the variability of performance metrics across specific partitions of the data and evaluate the generalizability and robustness of our models.

%% file: tables/patient_table.tex
\begin{table*}[t]
    \centering
    \caption{Patient characteristics table separated by cross-validation ($k^{th}$) fold.}
    \begin{tabular}{lccccc}
        \toprule
        $k^{th}$-fold                                & $k=1$       & $k=2$       & $k=3$       & $k=4$       & $k=5$       \\
        \hline
        Patients                                     & 175         & 175         & 175         & 175         & 175         \\
        Female                                       & 73          & 55          & 79          & 67          & 79          \\
        Male                                         & 102         & 120         & 96          & 108         & 96          \\
        Age                                          & 65 $\pm$ 11 & 64 $\pm$ 10 & 64 $\pm$ 10 & 62 $\pm$ 10 & 63 $\pm$ 11 \\
        Stem Cell Transplantees                      & 97          & 108         & 87          & 99          & 103         \\
        \hline
        \multicolumn{6}{l}{\textbf{ISS}} \\
        1                                            & 56          & 69          & 53          & 63          & 58          \\
        2                                            & 70          & 56          & 61          & 62          & 64          \\
        3                                            & 46          & 47          & 59          & 48          & 51          \\
        \hline
        \multicolumn{6}{l}{\textbf{Treatment Classification}} \\
        Bortezomib-based                             & 39          & 33          & 42          & 33          & 40          \\
        Carfilzomib-based                            & 2           & 0           & 2           & 0           & 0           \\
        IMIDs-based                                  & 11          & 14          & 6           & 8           & 10          \\
        combined IMIDs/carfilzomib-based             & 11          & 11          & 9           & 12          & 8           \\
        combined bortezomib/IMIDs-based              & 107         & 104         & 111         & 112         & 107         \\
        combined bortezomib/IMIDs/carfilzomib-based  & 5           & 12          & 5           & 9           & 9           \\
        combined bortezomib/carfilzomib-based        & 0           & 1           & 0           & 0           & 1           \\
        combined daratumumab/IMIDs/carfilzomib-based & 0           & 0           & 0           & 1           & 0           \\
        \hline
        Total Visits                                 & 3435        & 3269        & 3281        & 3483        & 3307        \\
        Total Progression Events                     & 240         & 253         & 197         & 255         & 292         \\
        \hline
        \multicolumn{6}{l}{\textbf{Lab Value Counts}} \\
        Hemoglobin                                   & 3056        & 2902        & 2806        & 3112        & 2876        \\
        Calcium                                      & 2992        & 2847        & 2748        & 3043        & 2799        \\
        Creatinine                                   & 3017        & 2863        & 2778        & 3066        & 2822        \\
        LDH                                          & 1999        & 2006        & 1882        & 1913        & 1951        \\
        Albumin                                      & 2929        & 2784        & 2711        & 3009        & 2756        \\
        $\beta$-2-Microglobulin                      & 1061        & 1028        & 850         & 791         & 977         \\
        M-Protein                                    & 2622        & 2585        & 2379        & 2692        & 2447        \\
        SFL-$\lambda$                                & 2806        & 2605        & 2503        & 2826        & 2630        \\
        SFL-$\kappa$                                 & 2818        & 2612        & 2510        & 2834        & 2625        \\
        WBC                                          & 3057        & 2904        & 2802        & 3108        & 2877        \\
        \hline
    \end{tabular}\label{tab:patient_table}
\end{table*}

%% file: sections/declarations/data_availability.tex
The CoMMpass data is available upon registration in the Multiple Myeloma Research Foundation (MMRF) Researcher Gateway at \url{https://research.themmrf.org}

%% file: sections/declarations/code_availability.tex
We hosted a \href{https://github.com/maximilianferle/predicting-progression-events-in-mm-from-routine-blood-work}{public GitHub repository} with all code used to produce the results of this study.

%% file: sections/declarations/acknowledgements.tex
The authors acknowledge the financial support by the Federal Ministry of Education and Research of Germany and by Sächsisches Staatsministerium für Wissenschaft, Kultur und Tourismus in the programme Center of Excellence for AI-research "Center for Scalable Data Analytics and Artificial Intelligence Dresden/Leipzig", project identification number: ScaDS.AI

This work was partially funded by grants from the International Myeloma Society (IMS Research Grant 2023), German Research Foundation (SPP µbone) and EU HORIZON (CERTAINTY).
The CERTAINTY project is funded by the European Union (Grant Agreement 101136379).
Views and opinions expressed are however those of the authors only and do not necessarily reflect those of the European Union.
Neither the European Union nor the granting authority can be held responsible for them.

The data used in this study were generated as part of the MMRF Personalized Medicine Initiatives (https://research.themmrf.org and https://themmrf.org).
We thank the MMRF for technical support and for facilitating access to their data.

%% file: sections/declarations/author_contributions.tex
MF conceptualized the analyses, designed and implemented the models, analysed the data, interpreted the results, designed figures and took the lead in writing the manuscript.

NG, MK, KR, AO and MM contributed to conceptualizing the analyses.

UP, TN, KR, AO and MM supervised the project.

KR, AO and MM devised the main conceptual ideas.

All authors discussed the results and contributed to writing the manuscript.

%% file: sections/declarations/competing_interests.tex
UP received financial support from
Syros Pharmaceuticals Inc.,
Silence Therapeutics GmbH,
Celgene GmbH,
Takeda Pharma Vertrieb GmbH,
Fibrogen Inc.,
Servier Deutschland GmbH,
Roche Deutschland Holding GmbH,
Merck KGaA,
Amgen Inc.,
Novartis Pharma GmbH,
Curis Inc.,
Janssen Pharmaceuticals Inc.,
Jazz Pharmaceuticals Inc.,
BeiGene Germany GmbH,
Geron Inc. and
Bristol-Myers Squibb GmbH.

AO received financial support from
Janssen-Cilag GmbH.

MM received financial support from
SpringWorks Therapeutics Inc.,
Janssen-Cilag GmbH and
Roche Deutschland Holding GmbH.

%% file: sections/supplementals.tex
\begin{center}
    \textbf{\large Supplementary Material}
\end{center}

\setcounter{equation}{0}
\setcounter{figure}{0}
\setcounter{table}{0}
\renewcommand{\theequation}{S\arabic{equation}}
\renewcommand{\thefigure}{S\arabic{figure}}
\renewcommand{\thetable}{S\arabic{table}}

\begin{figure*}[t]
    \includegraphics[width=\linewidth]{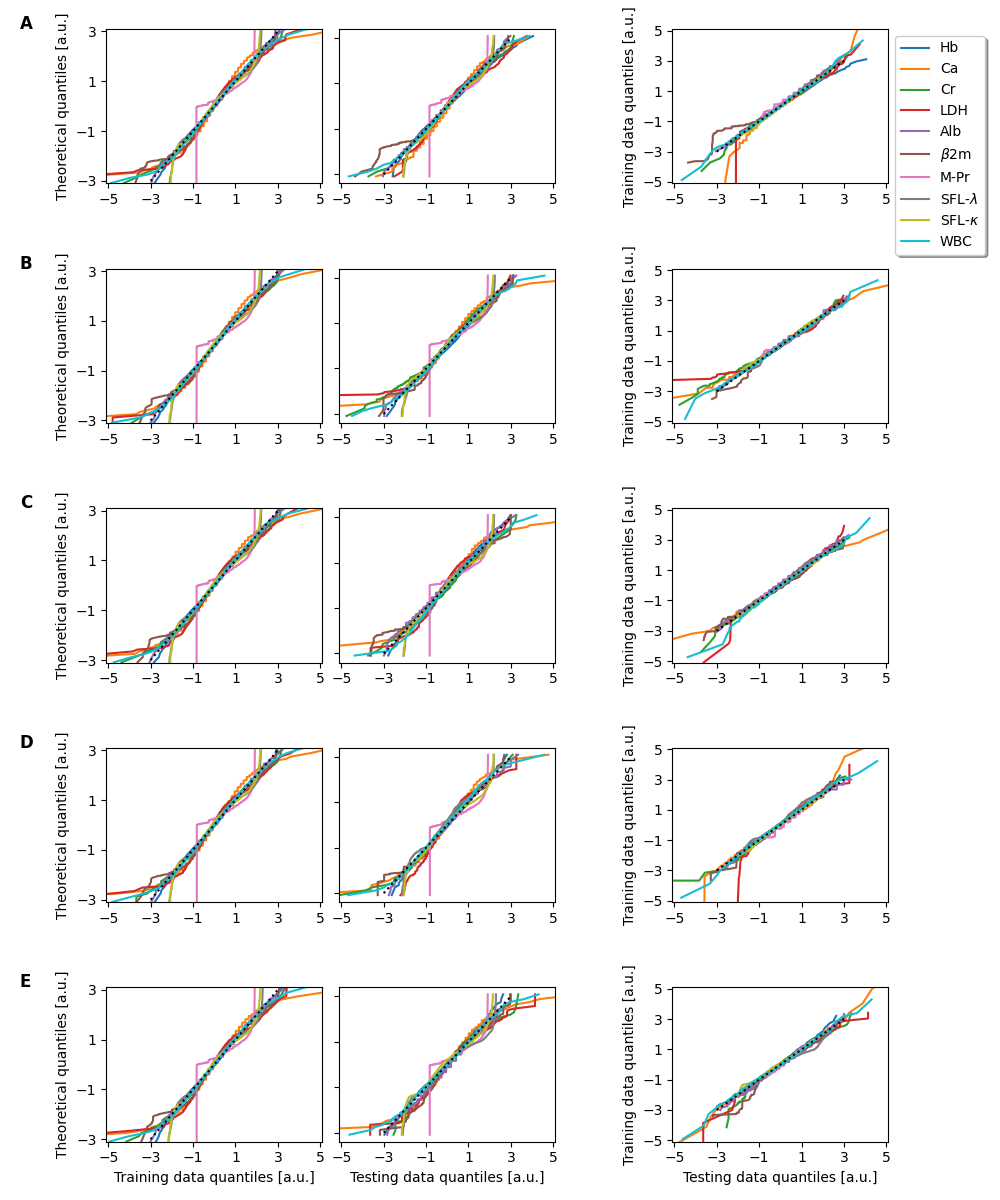}
    \caption{QQ-plots of the training and testing data after power transformation for $k=1$ (\textbf{A}), $k=2$ (\textbf{B}), $k=3$ (\textbf{C}), $k=4$ (\textbf{D}), $k=5$ (\textbf{E}). Theoretical quantiles indicate quantalies of a normal distribution. Dotted lines indicate the range of the central 99.9\% of the data.}
    \label{fig:qq}
\end{figure*}

\input{tables/r}
\input{tables/r_steps}

%% file: tables/r.tex
\begin{table}[h]
    \caption{Pearson's $r$ of absolute values}
    \centerline{
        \begin{tabular}{l|c|c|c|c|c}
            \toprule
            & \bf 3 months    & \bf 6 months    & \bf 9 months    & \bf 12 months   & \bf 15 months   \\
            \hline
            \bf Hb                      & 0.82 $\pm$ 0.02 & 0.77 $\pm$ 0.02 & 0.72 $\pm$ 0.02 & 0.67 $\pm$ 0.03 & 0.63 $\pm$ 0.03 \\
            \hline
            \bf Ca                      & 0.52 $\pm$ 0.09 & 0.44 $\pm$ 0.08 & 0.40 $\pm$ 0.08 & 0.37 $\pm$ 0.08 & 0.34 $\pm$ 0.08 \\
            \hline
            \bf Cr                      & 0.89 $\pm$ 0.01 & 0.87 $\pm$ 0.01 & 0.85 $\pm$ 0.01 & 0.83 $\pm$ 0.01 & 0.81 $\pm$ 0.01 \\
            \hline
            \bf LDH                     & 0.82 $\pm$ 0.02 & 0.76 $\pm$ 0.02 & 0.71 $\pm$ 0.03 & 0.68 $\pm$ 0.03 & 0.64 $\pm$ 0.03 \\
            \hline
            \bf Alb                     & 0.70 $\pm$ 0.02 & 0.63 $\pm$ 0.03 & 0.58 $\pm$ 0.03 & 0.55 $\pm$ 0.03 & 0.52 $\pm$ 0.03 \\
            \hline
            \bf $\beta$-2-Microglobulin & 0.92 $\pm$ 0.02 & 0.88 $\pm$ 0.03 & 0.84 $\pm$ 0.03 & 0.81 $\pm$ 0.03 & 0.77 $\pm$ 0.03 \\
            \hline
            \bf M-Pr                    & 0.82 $\pm$ 0.01 & 0.72 $\pm$ 0.02 & 0.64 $\pm$ 0.03 & 0.58 $\pm$ 0.03 & 0.53 $\pm$ 0.03 \\
            \hline
            \bf SFL-$\lambda$           & 0.82 $\pm$ 0.01 & 0.75 $\pm$ 0.02 & 0.70 $\pm$ 0.02 & 0.67 $\pm$ 0.03 & 0.63 $\pm$ 0.03 \\
            \hline
            \bf SFL-$\kappa$            & 0.84 $\pm$ 0.01 & 0.76 $\pm$ 0.01 & 0.71 $\pm$ 0.02 & 0.66 $\pm$ 0.03 & 0.62 $\pm$ 0.04 \\
            \hline
            \bf WBC                     & 0.68 $\pm$ 0.01 & 0.63 $\pm$ 0.02 & 0.59 $\pm$ 0.02 & 0.55 $\pm$ 0.02 & 0.52 $\pm$ 0.02 \\
        \end{tabular}\label{tab:abs_correlations}
    }
\end{table}

%% file: tables/r_steps.tex
\begin{table}[h]
    \caption{Pearson's $r$ of differences to last observation}
    \centerline{
        \begin{tabular}{l|c|c|c|c|c}
            \toprule
            & \textbf{3 months} & \textbf{6 months} & \textbf{9 months} & \textbf{12 months} & \textbf{15 months} \\
            \hline
            \textbf{Hb}                      & $0.42 \pm 0.03$   & 0.44 $\pm$ 0.03   & 0.45 $\pm$ 0.03   & 0.44 $\pm$ 0.03    & 0.44 $\pm$ 0.03    \\
            \hline
            \textbf{Ca}                      & 0.57 $\pm$ 0.05   & 0.60 $\pm$ 0.05   & 0.61 $\pm$ 0.05   & 0.62 $\pm$ 0.05    & 0.62 $\pm$ 0.04    \\
            \hline
            \textbf{Cr}                      & 0.45 $\pm$ 0.04   & 0.47 $\pm$ 0.04   & 0.46 $\pm$ 0.04   & 0.46 $\pm$ 0.04    & 0.45 $\pm$ 0.04    \\
            \hline
            \textbf{LDH}                     & 0.35 $\pm$ 0.07   & 0.39 $\pm$ 0.07   & 0.40 $\pm$ 0.06   & 0.40 $\pm$ 0.04    & 0.41 $\pm$ 0.05    \\
            \hline
            \textbf{Alb}                     & 0.48 $\pm$ 0.02   & 0.52 $\pm$ 0.02   & 0.53 $\pm$ 0.02   & 0.53 $\pm$ 0.02    & 0.54 $\pm$ 0.02    \\
            \hline
            \textbf{$\beta$-2-Microglobulin} & 0.26 $\pm$ 0.05   & 0.25 $\pm$ 0.03   & 0.25 $\pm$ 0.03   & 0.26 $\pm$ 0.02 & 0.26 $\pm$ 0.03 \\
            \hline
            \textbf{M-Pr}                    & 0.38 $\pm$ 0.02   & 0.44 $\pm$ 0.01   & 0.48 $\pm$ 0.02   & 0.50 $\pm$ 0.01    & 0.51 $\pm$ 0.01    \\
            \hline
            \textbf{SFL-$\lambda$}           & 0.40 $\pm$ 0.02   & 0.47 $\pm$ 0.02   & 0.49 $\pm$ 0.02   & 0.49 $\pm$ 0.02    & 0.48 $\pm$ 0.03    \\
            \hline
            \textbf{SFL-$\kappa$}            & 0.37 $\pm$ 0.03   & 0.44 $\pm$ 0.01   & 0.47 $\pm$ 0.01   & 0.49 $\pm$ 0.01    & 0.49 $\pm$ 0.02    \\
            \hline
            \textbf{WBC}                     & 0.53 $\pm$ 0.02   & 0.56 $\pm$ 0.02   & 0.57 $\pm$ 0.02   & 0.57 $\pm$ 0.02    & 0.56 $\pm$ 0.02    \\
        \end{tabular}
        \label{tab:step_correlations}
    }
\end{table}